\documentclass[aps,prl,twocolumn,showpacs,superscriptaddress,groupedaddress]{revtex4}  
\usepackage{graphicx}  
\usepackage{dcolumn}   
\usepackage{bm}        
\usepackage{amssymb}   
\usepackage{cases}
\usepackage{subfigure}
\begin{document}

\widetext

  \centerline{\em D\O\ INTERNAL DOCUMENT -- NOT FOR PUBLIC
DISTRIBUTION}


\title{Universal Einstein Relation Model in Disordered Organic Semiconductors under Quasi-equilibrium }
\author{Ling Li, Nianduan Lu, and Ming Liu }
\email{lingli@ime.ac.cn, liuming@ime.ac.cn}
\altaffiliation{Institute of Microelectronics, Chinese Academy of
Sciences, Beijing, 100029, China}

\date{\today}

\begin{abstract}
It is still under debate whether  the classical Einstein relation in
disordered organic semiconductors is valid. We investigated Einstein
relation in disordered organic semiconductors theoretically. The
results show that, the classic Einstein relation deviate
dramatically with disorder and electric field , even in the case of
thermal equilibrium.
\end{abstract}

\maketitle


Electron transfer and transport between organic molecules are the
fundamental issue of disordered organic semiconductors. Unlike
nearly perfect crystals, charge transport in amorphous and
polycrystalline thin films is dominated by various kinds of defects.
charge transport can be described in terms of variable- range
hopping, where the charges hop from site defect to site defect, the
probability of each hop depending on the site- energies and the
hopping distance. These physical properties deviate significantly
from classical semiconductor models. Therefore, the vality of the
Einstein relation to disordered semiconductors has been a matter of
intensive research. General speaking, the Einstein relation is the
relation between two fundamental transport parameters, the diffusion
coefficient $D$ and the mobility $\mu$ of charge carriers, which
reads as \cite{einstein}
\begin{equation}
\frac{D}{\mu}=\frac{k_BT}{q}
\end{equation}
with $k_B$ the Boltzmann constant, $T$ the temperature, and $q$ the
elementary charge. Numerous theoretical and experimental studies
\cite{basslereinstein1,basslereinstein2,basslereinstein3,basslereinstein4}
concluded that the Einstein relation is violated under
non-equilibrium conditions, mainly due to the electric field
dependence of the diffusivity being larger than the field response
of the mobility.\\
However, because of large disorder, charge transport in an organic
molecular crystal, being a sequence of transfers of an excess charge
on one molecule to one of its neighboring molecules, is a relatively
slow process and the time for carriers to equilibrate can be very
large, so that non- equilibrium transport is maybe not uncommon in
disordered semiconductors \cite{paul1,paul2}. Under
quasi-equilibrium conditions, it has been proposed that, the more
general Einstein relation should be written as
\cite{tessler1,tessler2,tessler3}
\begin{equation}
\frac{D}{\mu}=\frac{p}{q\partial p/\partial E_F}.
\end{equation}
Where $E_F$ is the Fermi-level and $p$ is the carrier concentration
relating to the density of states (DOS) as
\begin{equation}
p=\int\frac{g\left(E\right)}{1+\exp\left(\frac{E-E_F}{k_BT}\right)}dE.
\end{equation}
As a result, the Einstein relation becomes charge- density- and
temperature-dependent. However, the Einstein relation, defined as
equation (2), is till under suspicion. Even though such a derivation
is commonly accepted, Baranovskii et al. \cite{basslereinstein3,be5}
mentioned that equation (2) is valid only when $\mu$ and $D$ are
considered independent of energy, which is not the case for hopping
transport or relaxation in an exponential band tail.  Moreover, the
detailed hopping transport information \cite{bassler1,bassler2}, for
example, electric field and lattice spacing has
never been addressed in equation (2). \\
In this letter, we present here a universal model for Einstein
relation in disordered organic semiconductors under
quasi-equilibrium condition. The effect of the electric field,
temperature, and carrier concentration on the Einstein relation are
well addressed.

 \textit{Model}.---In general, the basis for models describing the charge
 transport in disordered semiconductors is Miller-Abrahams expressions
\cite{miller}, where the hopping transport takes place via tunneling
between an initial state $i$ and a target state $j$.The tunneling
process is described by

\begin{equation}
  \nu=\nu_0\exp\left(-u\right)=\nu_0\left\{
   \begin{array}{c}
   \exp\left(-2\alpha
R_{ij}-\frac{E_j-E_i}{k_BT}\right),  E_i>E_j\\
  \left(-2\alpha R_{ij}\right).  \qquad\qquad\qquad E_i<E_j\\
   \end{array}
  \right.
  \end{equation}
Here, $\nu_0$ is the attempt-to-jump frequency, $R_{ij}$ is the
hopping distance, $u$ is the hopping range
\cite{apsley,arkhipov2},$E_i$ and $E_j$ are the energies at sites
$i$ and $j$, respectively, and $\alpha$ is the inverse localized
length. However, in real organic semiconductor systems, when an
electric field $F$ exists, this electric field will lower the
Coulomb barrier, which leads to a reduction in the thermal
activation energies, and the hopping range with normalized energy
($\epsilon=E/k_BT$) can therefore be rewritten as \cite{apsley,li3}
\begin{equation}
 u=\left\{
   \begin{array}{c}
 2\alpha\left(1+\beta\cos\theta\right)
R_{ij}+\epsilon_j-\epsilon_i,  \epsilon_j>\epsilon_i-\beta\cos\theta\\
  2\alpha R_{ij}. \qquad\quad \qquad\qquad\qquad \epsilon_j<\epsilon_i-\beta\cos\theta\\
   \end{array}
  \right.
\end{equation}
where $\beta=Fe/{2\alpha k_BT}$ and $\theta$ is the angle between
$R_{ij}$ and the electric field ranging from $0$ to $\pi$. For a
site with energy $\epsilon_i$ in the hopping space, the most
probable hop for a carrier on this site is to an empty site at a
range $u$, for which it needs the minimum energy. The conduction is
a result of a long sequence of hops through this hopping space. For
simplicity, one-dimensional charge transport is taken at first. In
this situation, the average hopping range $R_{nn}$ can be obtained
following the approach used our previous work \cite{li3}, the
average hopping distance along the electric field, $\bar{x}_f$ is
given as

\begin{equation}
\bar{x}_f=\frac{I_1+I_2}{I_3+I_4}
\end{equation}
where
\begin{eqnarray*}
I_1=\sum\nolimits_{\pm}\int^{\epsilon_i+R_{nn}}_{\epsilon_i\pm
R_{nn}}g\left(\epsilon\right)\left(1-f\left(\epsilon,\epsilon_F\right)\right)\left[\frac{R_{nn}-\epsilon_i+\epsilon}{1\pm\beta}\right]x
d\epsilon
\end{eqnarray*}

\begin{eqnarray*}
I_2=\sum\nolimits_{\pm}\int^{\epsilon_i+R_{nn}}_{\epsilon_i\pm
R_{nn}}g\left(\epsilon\right)\left(1-f\left(\epsilon,\epsilon_F\right)\right)R_{nn}x
d\epsilon
\end{eqnarray*}

\begin{eqnarray*}
I_3=\sum\nolimits_{\pm}\int^{\epsilon_i+R_{nn}}_{\epsilon_i\pm
R_{nn}}g\left(\epsilon\right)\left(1-f\left(\epsilon,\epsilon_F\right)\right)\left[\frac{R_{nn}-\epsilon_i+\epsilon}{1\pm\beta}\right]
d\epsilon
\end{eqnarray*}
\begin{eqnarray*}
I_4=\sum\nolimits_{\pm}\int^{\epsilon_i+R_{nn}}_{\epsilon_i\pm
R_{nn}}g\left(\epsilon\right)\left(1-f\left(\epsilon,\epsilon_F\right)\right)R_{nn}
d\epsilon
\end{eqnarray*}

 The mobility at energy
$\epsilon_i$ is
\begin{equation}
\mu\left(\epsilon_i\right)=\lim_{t\rightarrow\infty}\frac{d\bar{x}_f}{Fdt}=\frac{\nu_0\bar{x}_f}{F}\exp\left(R_{nn}\right).
\end{equation}
Where $g\left(\epsilon\right)$ is the density of states, and
$f\left(\epsilon_i,\epsilon_F\right)=1/\left(1+\exp\left(\epsilon_i-\epsilon_F\right)\right)$
is the Fermi-Dirac distribution with $\epsilon_F$ denoting the Fermi
level. We take the Gaussian form of
 $g\left(\epsilon\right)=\frac{N_t}{\sqrt{2\pi}\sigma_0}\exp\left(-\frac{\epsilon^2}{2\sigma_0^2}\right)$
 \cite{bassler1}
as the DOS in the full manuscript, where $N_t$ is the number of
states per unit volume and $\sigma_0=\sigma'/kT$ indicates the width
of the DOS. $N_t=1\times 10^{28}m^{-3}$ is used in the full
manuscript as a typical value for the
 relevant organic semiconductor.

On the other hand, to calculate the diffusion constant $D$, we shall
use the following definitions
\begin{eqnarray*}
D\left(\epsilon_i\right)=\frac{1}{2}\lim_{t\rightarrow\infty}\frac{d}{dt}\left[\bar{x_f^2}-\bar{x_f}^2\right]=
\frac{\left[\bar{x_f^2}-\bar{x_f}^2\right]}{2}\\\times\nu_0R_{nn}\exp\left(-R_{nn}\right).
\end{eqnarray*}

The average the mean squared displacement of the carriers
$\bar{x_{f}^2}$ should be calculated as
\begin{equation}
\bar{x_f}^2=\frac{I'_1+I'_2}{I'_3+I'_4}
\end{equation}
where
\begin{eqnarray*}
I'_1=\sum\nolimits_{\pm}\int^{\epsilon_i+R_{nn}}_{\epsilon_i\pm
R_{nn}}g\left(\epsilon\right)\left(1-f\left(\epsilon,\epsilon_F\right)\right)\left[\frac{R_{nn}-\epsilon_i+\epsilon}{1\pm\beta}\right]^2x^2
d\epsilon
\end{eqnarray*}
\begin{eqnarray*}
I'_2=\sum\nolimits_{\pm}\int^{\epsilon_i+R_{nn}}_{\epsilon_i\pm
R_{nn}}g\left(\epsilon\right)\left(1-f\left(\epsilon,\epsilon_F\right)\right)R_{nn}^2x^2
d\epsilon
\end{eqnarray*}
\begin{eqnarray*}
I'_3=\sum\nolimits_{\pm}\int^{\epsilon_i+R_{nn}}_{\epsilon_i\pm
R_{nn}}g\left(\epsilon\right)\left(1-f\left(\epsilon,\epsilon_F\right)\right)\left[\frac{R_{nn}-\epsilon_i+\epsilon}{1\pm\beta}\right]^2
d\epsilon
\end{eqnarray*}
\begin{eqnarray*}
I'_4=\sum\nolimits_{\pm}\int^{\epsilon_i+R_{nn}}_{\epsilon_i\pm
R_{nn}}g\left(\epsilon\right)\left(1-f\left(\epsilon,\epsilon_F\right)\right)R_{nn}^2
d\epsilon
\end{eqnarray*}

\begin{figure}[h]
             \centering \scalebox{0.4}{\includegraphics{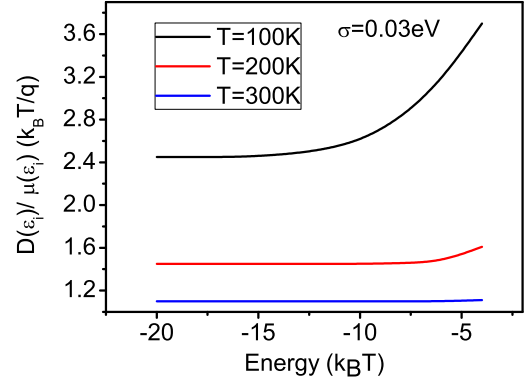}}
             \vspace*{-7pt}\caption{Dependence of Einstein relation on the energy for small material disorder at different temperatures. }
             \end{figure}

Based on the arguments above, the Einstein relation at energy
$\epsilon_i$ is given as
\begin{equation}
Ein(\epsilon_i)=\frac{D\left(\epsilon_i\right)}{\mu\left(\epsilon_i\right)}=\frac{\left[\bar{R_f}^2-\bar{R_f^2}\right]}{2\bar{x_f}}
\end{equation}
The overall Einstein relation in the hopping system can reasonably
formulated as

\begin{equation}
\frac{D}{\mu}=\frac{\int_{-\infty}^\infty
Ein\left(\epsilon_i\right)g\left(\epsilon_i\right)\left(1-f\left(\epsilon_i,\epsilon_F\right)\right)}{\int_{-\infty}^\infty
g\left(\epsilon_i\right)\left(1-f\left(\epsilon_i,\epsilon_F\right)\right)}
\end{equation}
To test the validity of Einstein relation, in Fig. 1, we firstly
plot the energy dependent of $Ein\left(\epsilon_i\right)$, using
equation (9). The parameters chosen here are typical ones for
organic semiconductors as: $F=1\times 10^5 V/m$, $\alpha^{-1}=1nm$,
$E_F=-20k_BT$,and $N_t=1\times 10^{28}$m$^{-3}$. Apparently, $eD/\mu
k_BT$  is actually energy independent at high temperature, and
approach to classic Einstein relation unity. However, this situation
changes dramatically with temperature decreasing. At low
temperature, the $eD/\mu k_BT$  deviate dramatically from unity and
will increase with starting energy.
 \begin{figure}[h]
             \centering \scalebox{0.4}{\includegraphics{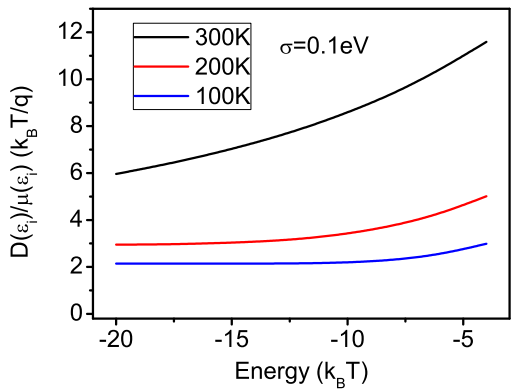}}
             \vspace*{-7pt}\caption {Dependence of $eD/\mu k_BT$ on the energy for larger material disorder at different temperatures.}
             \end{figure}
This deviation is more pronounced at large energy disorder, as
demonstrated clearly in Fig.2. In this situation, $eD/\mu k_BT$ is
as high as two times of unity, even at high temperature and deep
energy. Physically, since the hopping rate is actually energy
dependent, mobility or diffusion is therefore dependent on the
energy; Moreover, it has been pointed out that the energy dependent
transport in amorphous materials is more closely related to
diffusivity than mobility, $eD/\mu k_BT$ will deviate from unity
more rapidly at higher energies \cite{energy1,mott,mott2}. Figures 3
and 4 quantifies the deviation of overall  $eD/\mu k_BT$ (equation
10) from the classic Einstein relation by showing how $eD/\mu k_BT$
varies with the degree of disorder, temperature, and electric field.
The results are similar to the  $eD/\mu k_BT$ in non-equilibrium
transport, where the deviation increases with material disorder and
electric field, while decreases with temperature.

\begin{figure}[h]
             \centering \scalebox{0.4}{\includegraphics{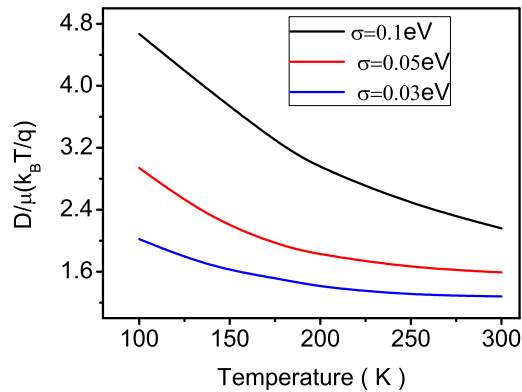}}
             \vspace*{-7pt}\caption{Dependence of $eD/\mu k_BT$ on temperature at different material disorder.}
             \end{figure}
Physical origin related to this deviation is similar to the
non-equilibrium as well, in energy or positional disorder
circumanal, the carrier predetermined paths defined by highest
exchange frequencies will govern carrier motion with certain
configurations being likely to override the tendency of a carrier to
follow the biasing field. The larger disorder, the more deviation.

Another interesting feature appears when  plotting $eD/\mu k_BT$  as
a function of the carrier concentration, as depicted in the inset of
Fig. 5. $eD/\mu k_BT$ here is found weakly dependent on carrier
concentration and decrease with concentration finally, which is in
contrast to the predictions based on equation (2)\cite{tessler2}. To
explain the above results we examine in Fig. 5 the carrier
concentration dependencies of both the diffusivity $D$ and mobility
$\mu$ of the charge carriers for $\sigma/k_BT=3.8$.
\begin{figure}[h]
             \centering \scalebox{0.4}{\includegraphics{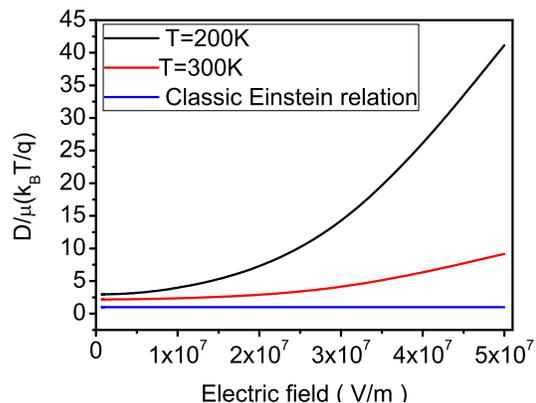}}
             \vspace*{-7pt}\caption{Dependence of $eD/\mu k_BT$ on the electric field  at different temperatures.}
             \end{figure}

Apparently,  both $D$ and $\mu$ increases with carrier
concentration. Remarkably, however, the diffusivity increases nearly
the same rapidly with carrier concentration as does $\mu$. The
constant deviations  is speculated from the temperature and material
disorder.

The question now coming out is whether these results are consistent
with the experimental data. The two available data for the
quasi-equilibrium regime are those of blom et al. \cite{blom1} and
keo et.al. \cite{german}. These data were obtained by measuring the
ideality factor in single carrier organic diode. It is well seen in
Fig. 3 that the relation between $D$ and $\mu$ differs from the
Einstein's formula dramatically, which agree with the observation in
\cite{german}. However, it has been argued that deviation of
Einstein relation in \cite{german}is an experimental artifact that
is caused by a too large leakage current
\cite{blom1}.Therefore, the comparison is not shown here.\\
It should be noted that, based on our calculation, the Einstein
relation is strongly dependent on material disorder, temperature,
and electric field, let $f\left(F,\sigma, T\right)=eD/\mu k_BT$. In
diffusion theory, drift and diffusion current $J$ of the M - i - n
diode reads as
\begin{equation}
J=n\mu\frac{\partial V}{\partial x}+D\frac{\partial n}{\partial x}.
\end{equation}
Following the classical derivation [], an intergrating factor
$\exp\left(-qV/f\left(T, F, \sigma\right)\right)$ is used to
integrate equation (11) over the $i$ layer ranging from $x=0$ to
$x=W$:
\begin{eqnarray}
J\int_0^W\exp\left(-\frac{qV}{f\left(T,F,\sigma\right)}\right)dx
\not=\mu
f\left(F,T,\sigma\right)\nonumber\\\times\left(n\exp\left(-\frac{qV}{f\left(T,F\sigma\right)}\right)\right)
\end{eqnarray}
Hence we conclude that the measured ideality factor can not be used
to prove or disprove the validity of the generalized Einstein
relation. This idea clearly calls for rigorous experimental work
which we hope this Letter will stimulate.

\begin{figure}[h]
             \centering \scalebox{0.4}{\includegraphics{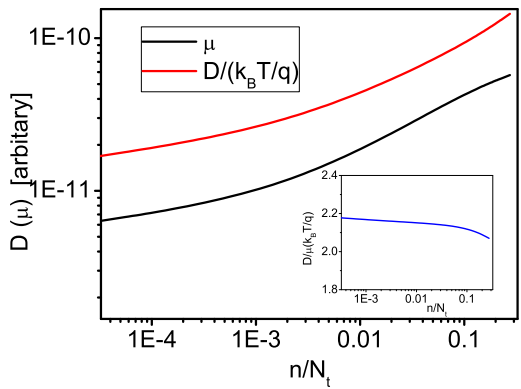}}
             \vspace*{-7pt}\caption{Concentration dependence of $D$($\mu$) on the carrier
             concentration. The inset shows the dependence of $eD/\mu k_BT$ on the carrier concentration.}
             \end{figure}

In conclusion, the universal model of the Einstein relation in
disordered organic semiconductors has been proposed here based on
variable range hopping theory. In contrast to earlier experimental
reports, deviation of $eD/\mu k_BT$ from unity is obtained.
Furthermore, it is found here  $eD/\mu k_BT$ is actually weakly
dependent on carrier concentration, which is in contradiction with
the generalized Einstein relation derived for a Gaussian DOS.
Furthermore, the valid of measuring ideality factor to check
Einstein relation is also discussed.

Financial support from NSFC (No. 60825403) and National 973 Program
2011CB808404 is acknowledged.

\end{document}